\documentclass[preprint,prc,aps,floatfix,showpacs,showkeys]{revtex4}
\usepackage{epsfig}
\usepackage{graphicx}
\begin{document}
\title{Failure of the Standard Coupled-Channels Method in Describing the Inelastic Reaction Data:
 On the Use of a New Shape for the Coupling Potential}
\author{S.Erturk\dag, I.Boztosun\ddag, Y.Kucuk\ddag\, M.Karakoc\ddag, S.Aydin\dag}
\affiliation{\dag\ Nigde University, Department of Physics,
Nigde-TURKEY. \\ \ddag\ Erciyes University, Department of Physics,
Kayseri-TURKEY.}

\begin{abstract} We present the failure of the standard
coupled-channels method in explaining the inelastic scattering
together with other observables such as elastic scattering,
excitation function and fusion data. We use both microscopic
double-folding and phenomenological deep potentials with shallow
imaginary components. We argue that the solution of the problems for
the inelastic scattering data is not related to the central nuclear
potential, but to the coupling potential between excited states. We
present that these problems can be addressed in a systematic way by
using a different shape for the coupling potential instead of the
usual one based on Taylor expansion.
\end{abstract}
\pacs{24.10.-i, 24.10.Eq, 24.10.-v, 24.10.+g}
\keywords{Coupled-Channels model, elastic/inelastic scattering,
$^{12}$C+$^{12}$C reaction} \maketitle
\section{Introduction} In this paper, we consider the $^{12}$C+$^{12}$C
reaction as a case study to point out the problems for the inelastic
scattering states which have so far remained unsolved, and to
address particularly the magnitude problem for the inelastic
scattering data. Theoretical calculations using  the
coupled-channels (CC) method fail to correctly predict the magnitude
of the single-2$^{+}$ and mutual-2$^{+}$ states data together with
the elastic scattering data. In order to get the magnitude right,
many futile theoretical attempts have been made for these states.
Previous theoretical works show that the shapes of the central real
potentials are actually correct, since they explain the elastic
scattering data and predict the resonances at the correct energies
with reasonable widths. It appears that the failure of the standard
methods is mostly related to the inelastic scattering data: The
magnitude of the theoretical cross-sections is much smaller than the
measured experimental data. In this paper, we make further
applications of a new coupling potential \cite{Boz1} in describing
the scattering observables of the $^{12}$C+$^{12}$C system.
\section{The Model}
For the spherical nuclei, the nuclear shape and also the shape of
the potential between projectile and target nuclei are characterized
by a constant radius $R$, which defines the distance of the center
of the nucleus from the surface. However, for a deformed nucleus,
the radius parameter is no longer constant but depends on the
angular location of the point ($R \rightarrow R(\theta,\phi)$). The
nucleus $^{12}$C we study in this paper is strongly deformed and its
collective excitation is taken into account by using the standard
deformation procedure based on the Taylor expansion. If the
interaction potential between two nuclei is taken to be
$U(r-R(\theta,\phi))$, the Taylor expansion about $R=R_{0}$ yields,
\begin{equation}
U(r-R)=U(r-R_{0})-\delta R \frac{\partial}{\partial r}U(r-R_{0})+
\frac{1}{2!}(\delta R)^{2} \frac{\partial^{2}}{\partial
r^{2}}U(r-R_{0})- \ldots \label{pot}
\end{equation}
Here, the first term is the usual diagonal optical potential that
describes only the elastic scattering and the second and third terms
are used to describe the inelastic scattering and to obtain the
coupling potentials for the single-2$^{+}$ and mutual-2$^{+}$
states. $\delta R$ in equation \ref{pot} is given as:
\begin{equation}
\delta R=R_{i}\sum_{lm}\beta_{lm}Y_{lm} (\theta,\phi)^{*}
\end{equation}
with $i$ as the projectile $a$ or the target $A$. $\beta_{lm}$ is
the deformation parameter and it is -0.6 for the $^{12}$C nucleus.

In the phenomenological analysis, the real nuclear potential has the
square of the Woods-Saxon shape, and the imaginary potential has the
Woods-Saxon volume shape. The parameters of the real and imaginary
parts are taken from ref. \cite{Boz1}.

For the microscopic analysis, the Nucleon-Nucleon double-folding
potential \cite{azab} is
\begin {equation}
 V_{DF}(R)=\int{\rho_{P}(r_{1})\rho_{T}(r_{1})V_{nn}(|\vec{R}+\vec{r}_{2}-\vec{r}_{1}|)d\vec{r}_{1}d\vec{r}_{2}}
\label {VDF}
\end  {equation}
where $\rho_{P}$ and $\rho_{T}$ are the nuclear matter distributions
for projectile and target nuclei respectively, and they are given by
\begin{equation}
\rho_{P}(r_{1})=\rho_{0P}(1+\gamma r_{1}^{2})exp(-\xi r_{1}^{2}) \hfill
\rho_{T}(r_{2})=\rho_{0T}[1+ exp((r_{2}-c)/a)]
\end{equation}
where $\rho_{0P}$=0.1644 fm$^{-3}$, $\gamma$=0.4988 fm$^{-2}$ and
$\xi$=0.3741 fm$^{-2}$ for projectile and $\rho_{0T}$=0.207
fm$^{-3}$, c=2.1545 fm, and a=0.425 fm for target nuclei.

\begin{figure}[ht]
\epsfxsize 13.5cm \centerline{\epsfbox{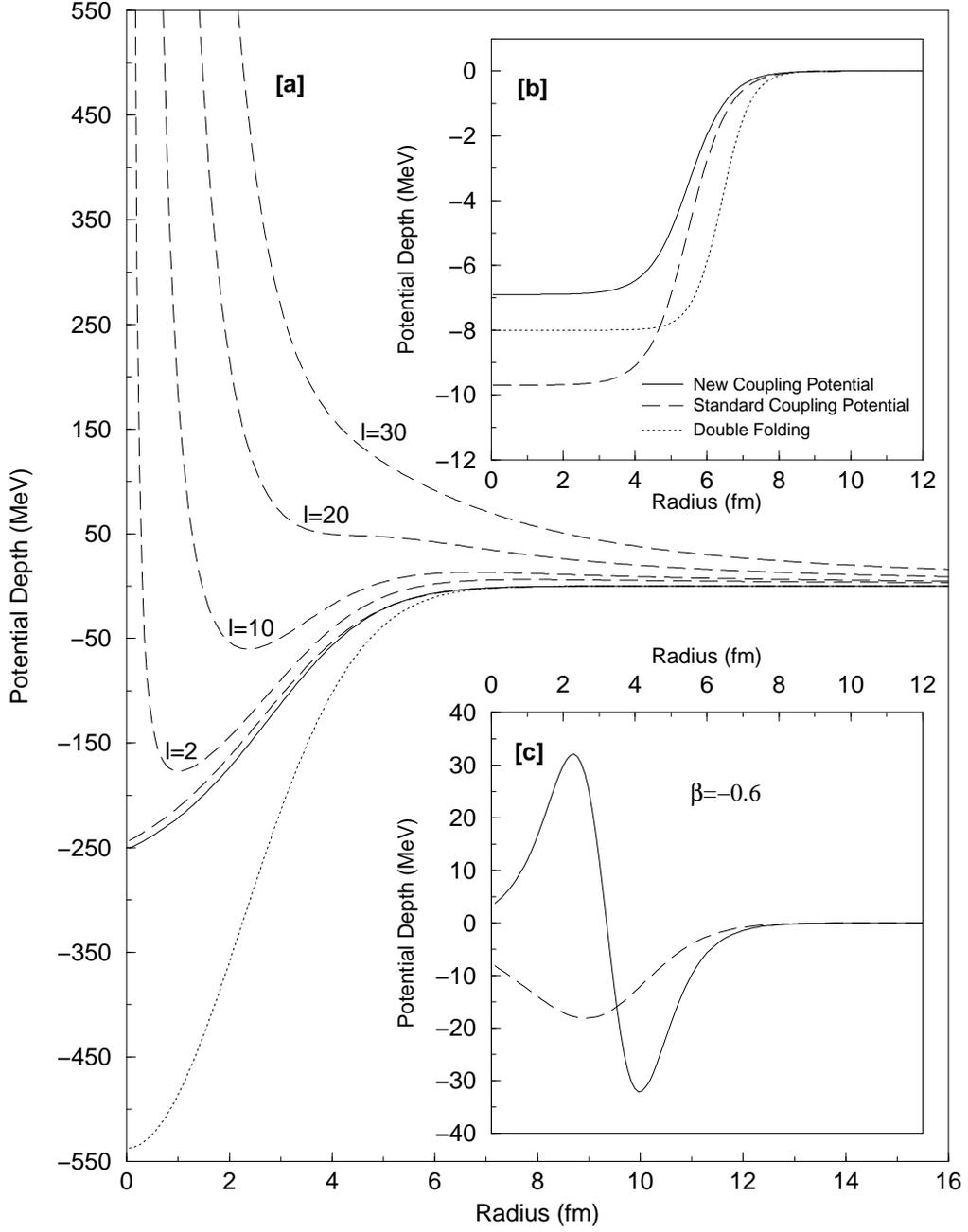}}
\caption{Potentials: [a] the DF and WS$^{2}$, [b] the imaginary and
[c] the standard WS$^{2}$ and new coupling potentials at
$E_{lab}$=74.2 MeV.}
 \label{comp}
\end{figure}

The M3Y Nucleon-Nucleon effective interaction is taken in the form
\begin {equation}
V_{nn}(r)=7999\frac{exp(-4r)}{4r}-2134\frac{exp(-2.5r)}{2.5r}+J_{00}(E)\delta(s)
\label{eff_nn}
\end{equation}
where $J_{00}(E)=276(1-0.005 \frac{E}{Ap})$. The real and imaginary
potentials are shown in figure \ref{comp} and the parameters are
given in table \ref{param} labelled as DF.

In the new coupled-channels model, we have replaced the usual first
derivative coupling potential by a second-derivative coupling
potential in Woods-Saxon form which is multiplied by the diffuseness
parameter ($a$) to normalize the units in the calculations. The
parameters are given in table \ref{param}.
\section{Results}
\begin{figure}[ht]
\epsfxsize 13.5cm \centerline{\epsfbox{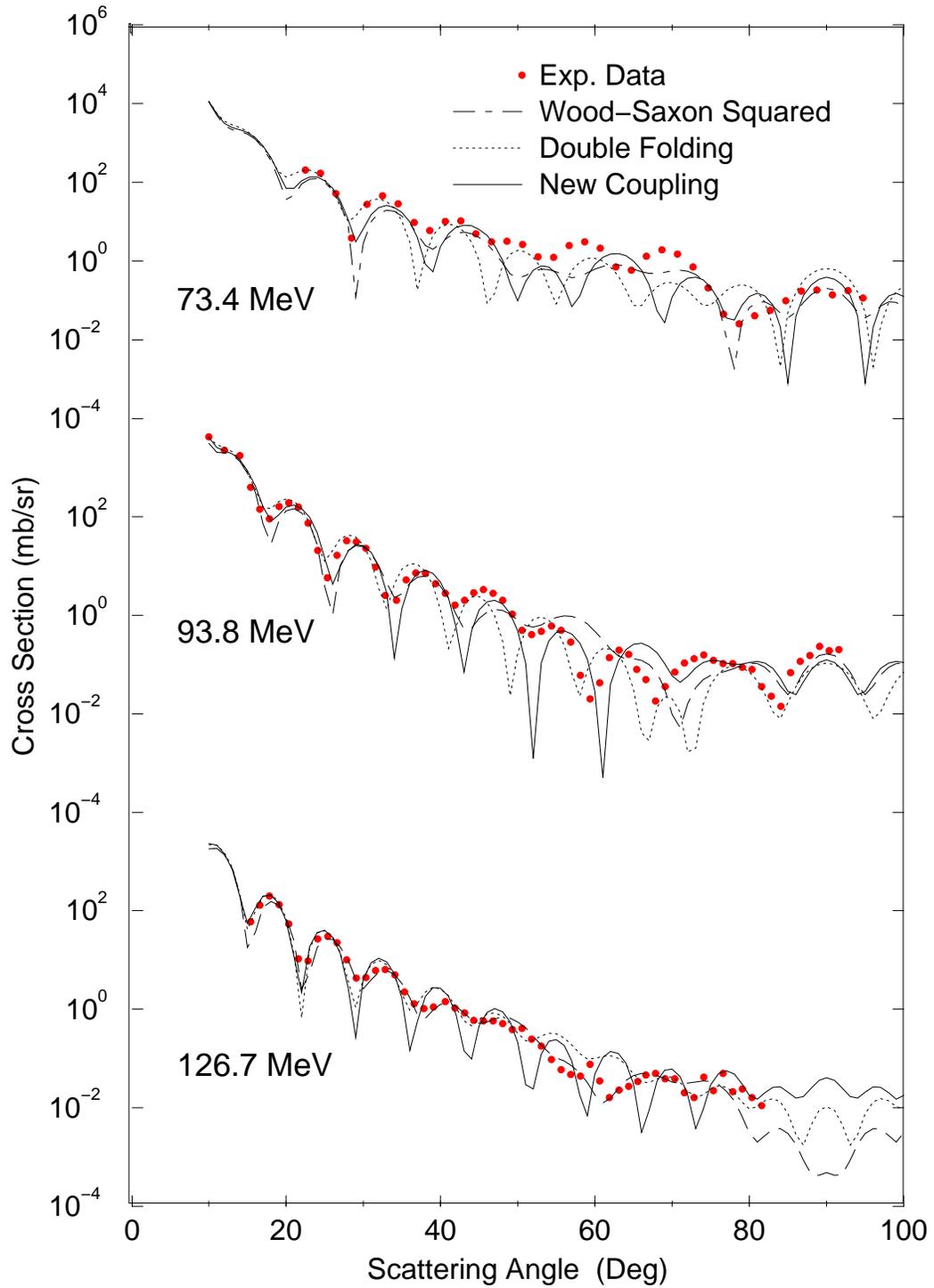}}\caption{Ground
state results: The dotted and long-dashed lines show the predictions
of DF and WS$^{2}$ potentials with standard coupling while the solid
lines show the results of the new coupling potential.}
 \label{ground}
\end{figure}

\begin{figure}[ht]
\epsfxsize 13.5cm \centerline{\epsfbox{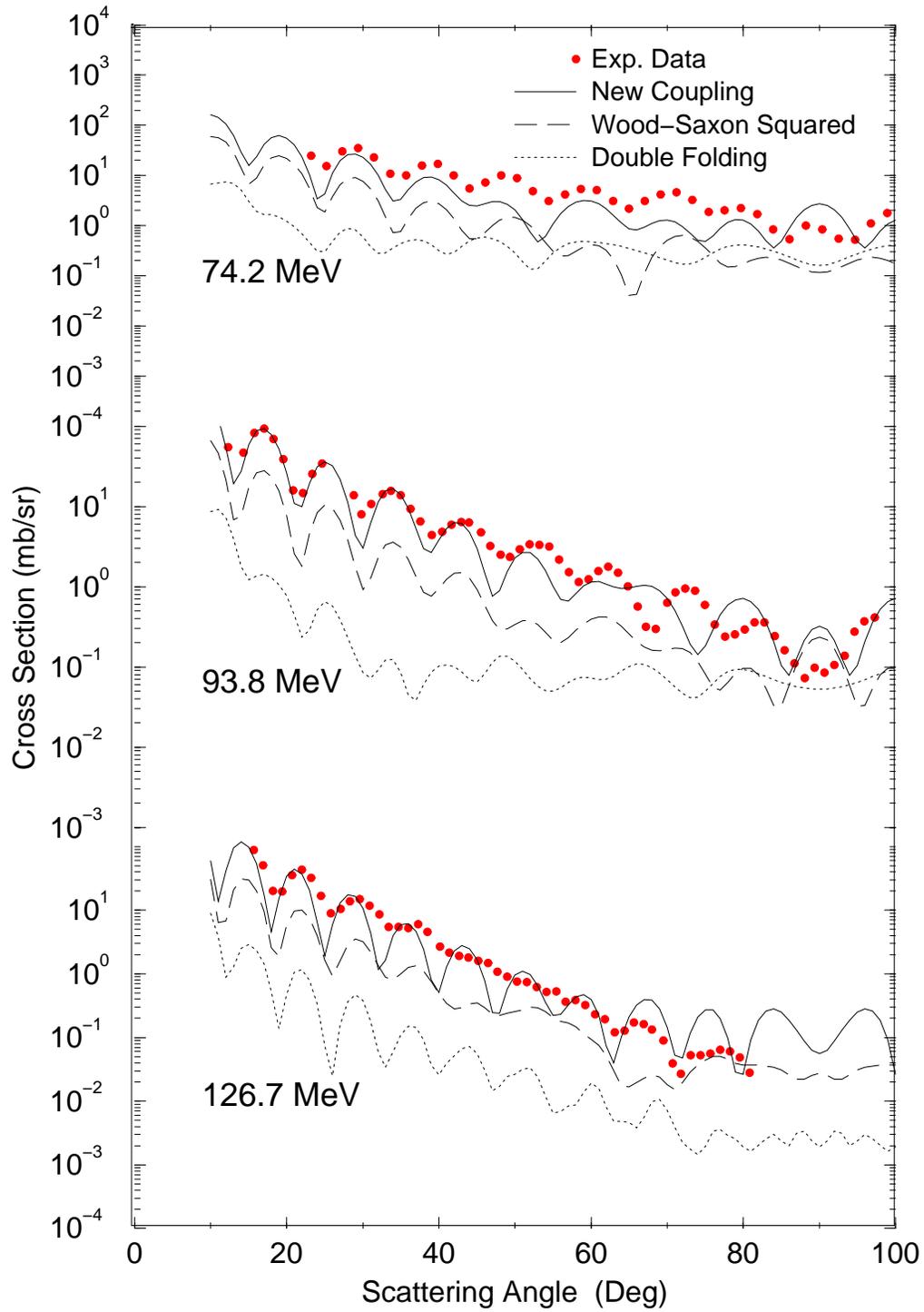}}
\caption{Single-2$^{+}$ state results. The legends are the same as
figure \ref{ground}.}
 \label{single}
\end{figure}

\begin{figure}[ht]
\epsfxsize 13.5cm \centerline{\epsfbox{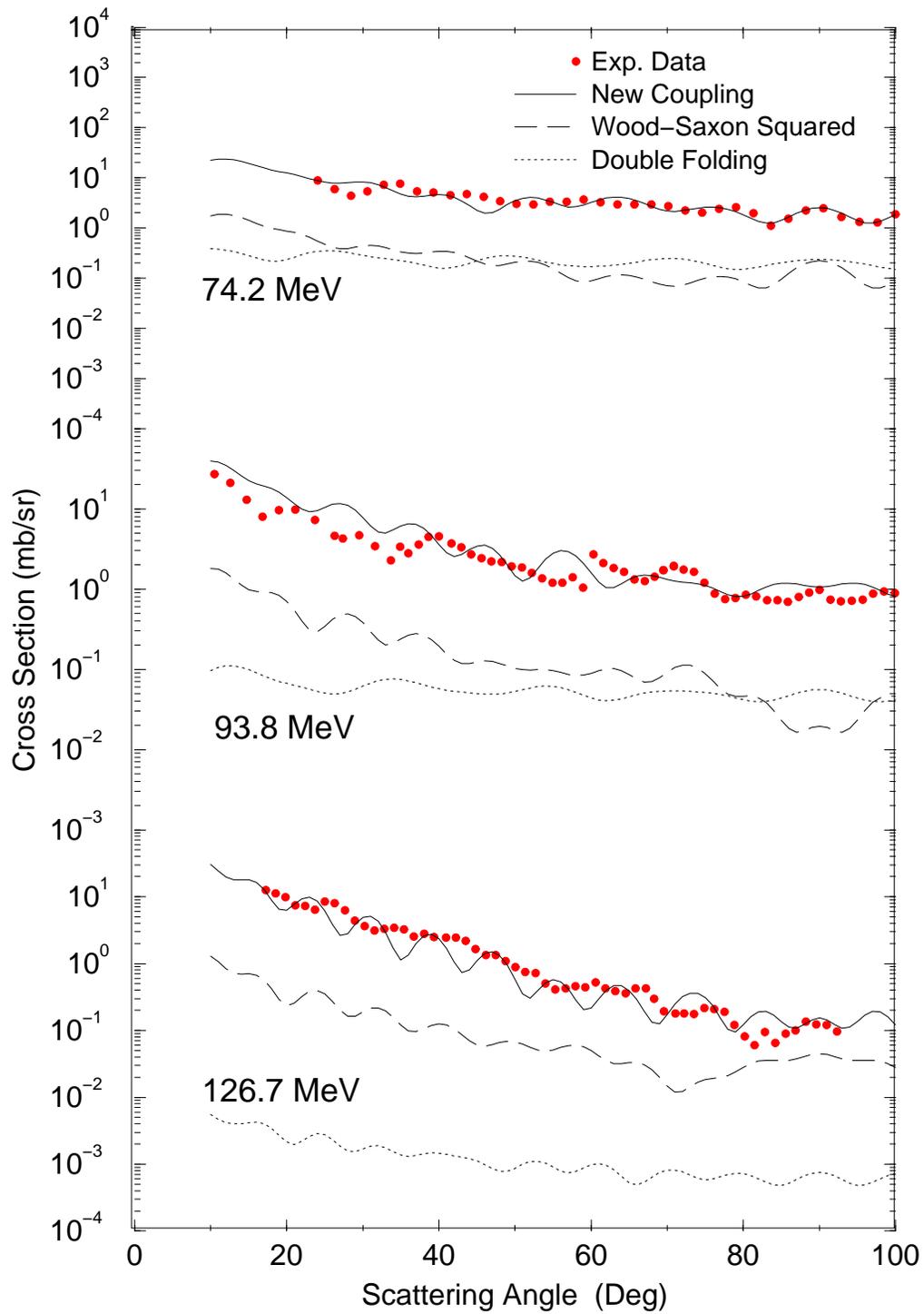}}
\caption{Mutual-2$^{+}$ state results. The legends are the same as
figure \ref{ground}.}
 \label{mutual}
\end{figure}

We have used both the phenomenological and microscopic
potentials to analyze the experimental data of the $^{12}$C+$^{12}$C
reaction at E$_{Lab}$=74.2 MeV, 93.8 MeV and 126.7 MeV. The
experimental data is taken from ref. \cite{Sto79}. The results of
our analyzes are displayed in Figure \ref{ground} for the ground,
\ref{single} for the single-2$^{+}$, and \ref{mutual} for the
mutual-2$^{+}$ states in comparison with experimental data. Both
double-folding and phenomenological potentials provide excellent
agreement with the experimental data for the ground state at all
energies and a good fit to the single-2$^{+}$ state data. However,
the mutual-2$^{+}$ state prediction is much smaller than the
measured one: The standard CC model using double-folding or
phenomenological potentials underestimates its magnitude by a factor
of 3 to 10 as it has been previously
observed~\cite{Boz1,Sto79,Fry97,Rae97,Sak99,Wol82}. Varying the
parameters and changing the shape of the real and imaginary
potentials do not provide a complete solution to the problems of
this reaction~\cite{Boz1}. In order to solve this problem we have
used  a new coupling potential. This potential has a
second-derivative of Woods-Saxon shape and it is compared in figure
\ref{comp} with the standard coupling potential. This new coupling
potential has a repulsive part at short distances and an attractive
part at large distances which is related to the orientation of two
$^{12}$C nuclei at short and large distances \cite{Boz1}. We have
been able to obtain excellent agreement with all the available
experimental data for the ground, single-2$^{+}$ and mutual-2$^{+}$
states by using this new coupling potential. The parameters are
shown in table~\ref{param}. This new approach solves the magnitude
problem of the mutual-2$^{+}$ state data, which has been an
outstanding problem with this reaction. The results for the ground,
single-2$^{+}$ and mutual-2$^{+}$ states are compared with the
standard ones in figures \ref{ground}, \ref{single} and
\ref{mutual}.
\section{Summary} In the present work, we have demonstrated that a
consistent solution could be obtained for the problems of the
$^{12}$C+$^{12}$C reaction over a wide energy range. However, we
achieve this by using a coupling potential which has a non-standard
form. Within the standard formalism, our findings using folding or
phenomenological potentials are in agreement with the previous works
\cite{Boz1,Sto79,Fry97,Rae97,Sak99,Wol82}. Although, within standard
approach, these potentials give excellent agreement with the
experimental data for the ground state, they are unable to provide a
consistent solution to the problems of the inelastic scattering
data. This work therefore clearly shows that the standard
deformation procedure based on Taylor expansion is inadequate in
describing such highly deformed nuclei. It is obvious that the
standard formalism should be questioned further.
\section*{Acknowledgments} This work is supported by the Turkish Science and Research
Council (T\"{U}B\.{I}TAK): Grant No: TBAG-2398 and Erciyes
University-Institute of Science: Grant no: FBT-04-15 and FBT-04-16.

\newpage
\begin{table}
\begin{center}
\begin{tabular}{lrcccrrrr} \hline \hline
     &$E_{lab}$(MeV)&N  &  V(MeV)  &r$_{v}$(fm)& a$_{v}$(fm)& W(MeV) &r$_{w}$(fm)& a$_{w}$(fm)
\\\hline
DF       & 74.2  & 1.47&   -  &  -   &  -    &  8.0& 1.40& 0.40\\
         & 93.8  & 1.20&   -  &  -   &  -    & 10.0& 1.40& 0.40\\
         &126.7  & 1.20&   -  &  -   &  -    & 23.0& 1.20& 0.50\\
         \hline
New      & 74.2  &  -  & 285.0&0.810 & 1.35  &  6.9& 1.20& 0.55\\
         & 93.8  &  -  & 285.0&0.760 & 1.35  &  9.5& 1.20& 0.55\\
         &126.7  &  -  & 285.0&0.800 & 1.35  & 15.0& 1.20& 0.55\\
\hline\hline
\end{tabular}
\end{center}
\caption{The parameters of the real and imaginary potentials. For
the new coupling potentials, the radius and  diffuseness 0.7 fm and
the depth is 215.0 MeV, 230 MeV and 245 MeV for $E_{lab}$=74.2 MeV,
93.8 MeV and 126.7 MeV.}
 \label{param}
\end{table}


\end{document}